\documentclass{article}
\usepackage{spconf,amsmath,graphicx}
\usepackage{algorithm}
\usepackage{algpseudocode}
\usepackage{amssymb}

\title{SELF-ADAPTIVE SOFT VOICE ACTIVITY DETECTION USING DEEP NEURAL NETWORKS FOR ROBUST SPEAKER VERIFICATION}
\name{Youngmoon Jung, Yeunju Choi, Hoirin Kim}
\address{School of Electrical Engineering, KAIST, Daejeon, South Korea}
\begin{document}
%
\maketitle
\begin{abstract}
Voice activity detection (VAD), which classifies frames as speech or non-speech, is an important module in many speech applications including speaker verification.
In this paper, we propose a novel method, called self-adaptive soft VAD, to incorporate a deep neural network (DNN)-based VAD into a deep speaker embedding system. 
The proposed method is a combination of the following two approaches.
The first approach is soft VAD, which performs a soft selection of frame-level features extracted from a speaker feature extractor.
The frame-level features are weighted by their corresponding speech posteriors estimated from the DNN-based VAD, and then aggregated to generate a speaker embedding.
The second approach is self-adaptive VAD, which fine-tunes the pre-trained VAD on the speaker verification data to reduce the domain mismatch.
Here, we introduce two unsupervised domain adaptation (DA) schemes, namely speech posterior-based DA (SP-DA) and joint learning-based DA (JL-DA). 
Experiments on a Korean speech database demonstrate that the verification performance is improved significantly in real-world environments by using self-adaptive soft VAD.

\end{abstract}
\begin{keywords}
speaker verification, voice activity detection, unsupervised domain adaptation, soft VAD 
\end{keywords}
\section{Introduction}
\label{sec:intro}

Speaker verification (SV) is the task of verifying a person's claimed identity based on his or her voice.
An important component of a practical SV system is voice activity detection (VAD), which detects the speech regions of an utterance which are the most effective for speaker discrimination.
For example, if too many non-speech segments are misclassified as speech and used in the training, then it can corrupt background models and hence significantly reduces the performance of SV systems.
On the other hand, during testing, if not enough speech segments are detected, then the SV algorithms will not be able to detect the speaker. 
For this reason, the VAD has played a vital role in robust SV systems from traditional Gaussian Mixture Model-Universal Background Model (GMM-UBM) and i-vector systems \cite{McLaren2015, Yamamoto2017} to recent deep speaker embedding systems \cite{Wang2018, Okabe2018, Zeinali2019, Tang2019}.

However, SV and VAD techniques have been largely developed independently of each other.
Research on the use of VAD in the SV context is surprisingly limited.
Most modern SV systems still use a traditional energy-based VAD, possibly due to its simplicity \cite{Wang2018, Okabe2018, Zeinali2019, Tang2019}.
In high signal-to-noise ratio (SNR) conditions, the energy-based VAD works reasonably well, while in low SNR environments, it produces unreliable speech frames \cite{Sahidullah2012, Yu2011}.
To deal with this problem, several deep neural network (DNN)-based VADs \cite{Bie2015, Jung2017,Jung2018,Fan2019} have been proposed and shown to give better results in low SNRs.
Therefore, it is more desirable to use DNN-based VADs instead of energy-based VADs for SV systems in real-world environments where background noise is always present and the SNR may not be high enough to apply the energy-based VAD. 

In this work, we propose an algorithm, self-adaptive soft VAD, to integrate the DNN-based VAD into the deep speaker embedding based SV system.
The proposed algorithm is a combination of two algorithms. The first algorithm is soft VAD, which was proposed in \cite{Wang2018}. 
Soft VAD generates frame-wise speech posteriors and integrates these probabilities directly into an SV system instead of making a hard decision based on a threshold, which is performed in general VAD. 
The frame-level features extracted from a speaker feature extractor are weighted by their corresponding speech posteriors estimated from the VAD model. Soft VAD can be combined with self-adaptive VAD and used to backpropagate the gradient of the loss of the speaker embedding network through the VAD model. In this way, the VAD can be adapted to the SV domain. Another advantage of using soft VAD is that it removes the need to determine the optimal threshold value to make a binary speech/non-speech decision.

The second algorithm for integrating VAD into the speaker verification system is self-adaptive VAD.
In a general setting, VAD and SV models are trained using different datasets.
Therefore, when we apply VAD for SV, the domain mismatch between training and test data can lead to a significant degradation in the performance of VAD. To reduce the domain mismatch, we propose two fine-tuning based unsupervised domain adaptation (DA) methods: speech posterior-based DA (SP-DA) and joint learning-based DA (JL-DA). 

For the SP-DA method, we fine-tune a pre-trained VAD on the SV data. This fine-tuning based domain adaptation method has been applied in other tasks \cite{Hoffman2014, Oquab2014, Bak2018}.
Here, the problem is that the SP-DA is a supervised method that requires VAD labels for the SV data, which is costly to obtain. Therefore, we need an unsupervised method that does not require any labeling information from the target domain. 
This is achieved by thresholding speech posteriors estimated from the VAD to generate ``reliable" labels for each utterance.
Then, the VAD is fine-tuned using labels generated by the VAD itself, and the process is repeated.

For the JL-DA method, we first integrate the VAD into the SV system through a soft VAD algorithm. As we already mentioned above, the gradient of the loss of the speaker embedding network is backpropagated through the VAD. Since the VAD process is partly guided by the loss of the speaker embedding network, the VAD would hopefully be able to produce higher posterior probabilities for frames which are more important for the SV task.
The self-adaptive VAD is then conducted by combining two domain adaptation approaches.

In this paper, we first review related prior works in Section 2.
Then section 3 presents our proposed method, self-adaptive soft VAD. The experimental setups and results are described in Section 4 and Section 5, respectively. We conclude this work in Section 6.

\section{PRIOR WORKS}
\label{sec:format}

There have been a few studies that have investigated the combination of VAD and SV. These studies can be divided into two main categories: soft VAD and self-adaptive VAD.
In this section, we review the two previous approaches, respectively.

\subsection{Soft VAD}
Soft VAD was first proposed in \cite{McLaren2015}. The purpose of their work is to improve the robustness of speaker recognition under mismatched train/test conditions by reducing the dependence of VAD on the tuned threshold. To achieve this, they directly integrated speech posteriors into a speaker recognition system instead of using VAD (or called hard VAD). 
Specifically, GMM-based VAD generates frame-wise speech posteriors and employs these posteriors in order to suppress the impact of the non-speech like frames on the speaker factors. The latter is achieved by weighting each frame with its speech posterior during the calculation of Baum-Welch statistics in the i-vector framework.
Soft VAD improves the generalization of speech/non-speech models to unseen conditions by removing the need to make binary speech/non-speech decisions based on a threshold. They demonstrated the benefits of soft VAD over hard VAD in severely mismatched conditions. 

In \cite{Yamamoto2017,Wang2018}, DNN-based VAD was employed for soft VAD. 
Yamamoto \textit{et al.} \cite{Yamamoto2017} applied the same soft VAD method for i-vector extraction. They used LSTM-based VAD instead of GMM-based VAD to produce a frame-wise speech posterior. Wang \textit{et al.} \cite{Wang2018} employed the same LSTM-based VAD to the deep speaker embedding network \cite{Snyder2017}. 
The authors combined attention weights in attentive statistics pooling \cite{Okabe2018} with speech posteriors estimated from the LSTM-based VAD. The product of the attention weight and speech posterior is multiplied with the corresponding frame-level speaker feature. In this paper, we employ the same soft VAD method to integrate the DNN-based VAD into the deep speaker embedding network and refer to this approach as ``attention-based soft VAD" to distinguish it from other soft VAD methods \cite{McLaren2015, Yamamoto2017}. More details will be discussed in Section 3.1.

\subsection{Self-adaptive VAD}
Kinnunen \textit{et al.} \cite{Kinnunen2013} proposed a vector quantization (VQ)-based self-adaptive VAD (VQ-VAD) for i-vector based SV system.
VQ-VAD shows better performance than energy-based VAD, especially in noisy conditions.
The main algorithm steps are as follows:
First, they extract mel-frequency cepstral coefficients (MFCC) features from the original noisy speech signal. 
Next, the spectral subtraction is applied to the noisy speech for speech enhancement. 
The purpose of this step is to merely increase the energy contrast between speech and non-speech.
Following this, the frames are sorted by their log-energy values in ascending order.
A fixed percentage of the lowest and highest energy frames (for instance, 10\% of all frames in each case) are assumed to correspond, respectively, to reliably-labeled non-speech and speech frames.
Using k-means (k = 16) clustering, GMMs of both speech and non-speech are trained taking the MFCCs corresponding to the lowest and highest energy frame indices. Finally, all the frames are labeled using the trained models, with an additional minimum-energy constraint.

Asbai \textit{et al.} \cite{Asbai2015} 
improved VQ-VAD by using maximum \textit{a posteriori} (MAP) adaptation.
They first create two UBMs for speech and non-speech models trained from long utterances. Then, an adaptation of UBMs to the short utterance of a speaker is performed via MAP adaptation. 


\section{PROPOSED APPROACH}
In this section, we introduce the proposed approach, self-adaptive soft VAD, which combines attention-based soft VAD and DNN-based self-adaptive VAD. 
We explain each of them in the following subsections.

\begin{figure}[t]
  \vspace{-0.8cm}
  \centerline{\includegraphics[width=8.8cm]{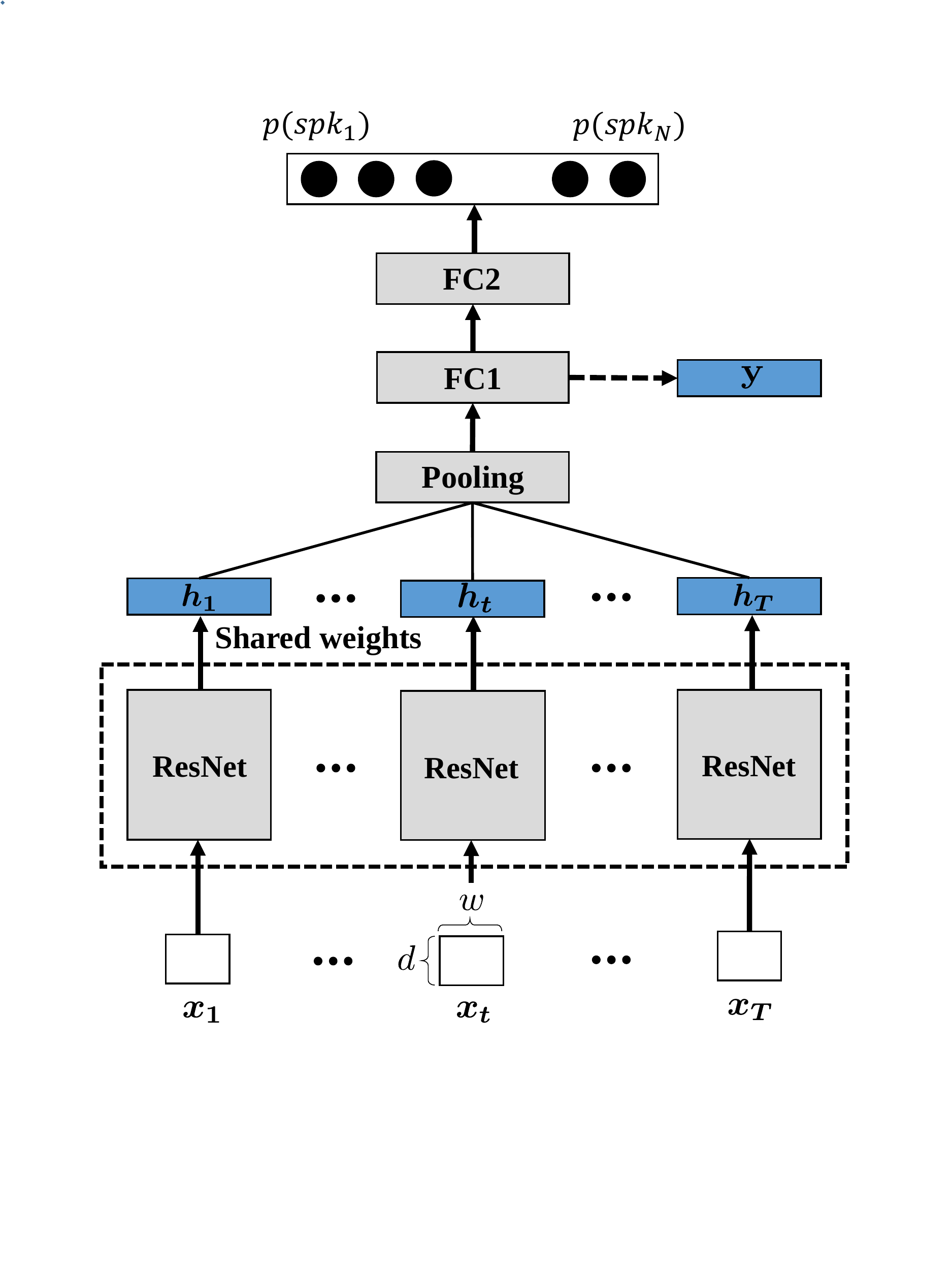}}
  \vspace{-2.4cm}
  \caption{ResNet-based deep speaker embedding system for extracting utterance-level speaker embeddings.}
  \label{SR_basic}
\end{figure}

\begin{table}[t]
\centering
\begin{footnotesize}
\renewcommand{\arraystretch}{1.2}
\vspace{-0.1cm}
\caption{The architecture of the frame-level feature extractor based on 34-layer ResNet \cite{He2016}. The input size is $d \times w$.}
\vspace{0.5cm}
\label{architecture}
\begin{tabular}{c|c|cll}
\cline{1-3}
stage & output size    & ResNet-34 \\ \cline{1-3}
conv1      & $d \times w \times 16$    & $7 \times 7, 16$, stride $1$                                              \\ \cline{1-3}
conv2       & $d \times w \times 16$    & $\left[ \begin{array}{cc} 3 \times 3, 16  \\ 3 \times 3, 16 \end{array}\right]$ $\times$ $3$ \\ \cline{1-3}
conv3       & $d/2 \times w/2 \times 32$  & $\left[ \begin{array}{cc} 3 \times 3, 32  \\ 3 \times 3, 32 \end{array}\right]$ $\times$ $4$ \\ \cline{1-3}
conv4       & $d/4 \times w/4 \times 64$ & $\left[ \begin{array}{cc} 3 \times 3, 64  \\ 3 \times 3, 64 \end{array}\right]$ $\times$ $6$ \\ \cline{1-3}
conv5       & $d/8 \times w/8 \times 128$  & $\left[ \begin{array}{cc} 3 \times 3, 128  \\ 3 \times 3, 128 \end{array}\right]$ $\times$ $3$ \\ \cline{1-3}
            & $1 \times 1 \times 128$  & global average pooling \\ \cline{1-3}
\end{tabular}
\end{footnotesize}

\end{table}

\subsection{Attention-based soft VAD with the SV system}
Before we explain the attention-based soft VAD, we first describe the deep speaker embedding system in detail.
Fig. \ref{SR_basic} illustrates the deep speaker embedding system used in this paper, which consists of three modules. The first module is the frame-level feature extractor which takes a sequence of acoustic features $\boldsymbol{x}_t$ and outputs corresponding speaker features $\boldsymbol{h}_t\,(t=1,\dotsm,T)$. In our system, ResNet \cite{He2016} is used as a feature extractor, which has been widely used in previous studies \cite{Li2017, Cai2018, Jung2019}. 
The architecture is described in Table \ref{architecture}. 
Here, 40-dimensional log Mel-filterbank (Fbank) features are used as acoustic features. The 11-frame context window is appended to form the $40 \times 11$ time-frequency feature maps for each frame (i.e., $d=40$ and $w=11$). 
The ResNet takes Fbank features of size $40 \times 11$ and outputs 128-dimensional frame-level features. 

The second block of the deep speaker embedding is a pooling layer that converts variable-length frame-level features into a fixed-dimensional vector. We apply self-attentive pooling \cite{Cai2018} that provides importance-weighted means of frame-level features, for which the importance is calculated by an attention mechanism. An attention model calculates a scalar score $e_{t}$ for the frame-level feature $\boldsymbol{h}_t$:
\begin{equation}
e_t = \boldsymbol{v}^Tf(\boldsymbol{Wh}_{t}+\boldsymbol{b})\,,
\end{equation}
where $f(\cdot)$ is a ReLU activation function. 
The score is normalized over all frames by a softmax function:
\begin{equation}
\alpha_t = \frac{exp(e_t)}{\sum_{i=1}^T exp(e_i)}\,.
\end{equation}
The normalized score $\alpha_t$ is then used as the weight in the pooling layer to calculate the following weighted mean vectors:
\begin{equation} \label{eq:att_weightedsum}
\boldsymbol{\widetilde{\mu}} = \sum_{t=1}^T \alpha_t \boldsymbol{h}_t\,.
\end{equation}

\begin{figure}[t]
  \vspace{-0.7cm}
  \centerline{\includegraphics[width=11.5cm]{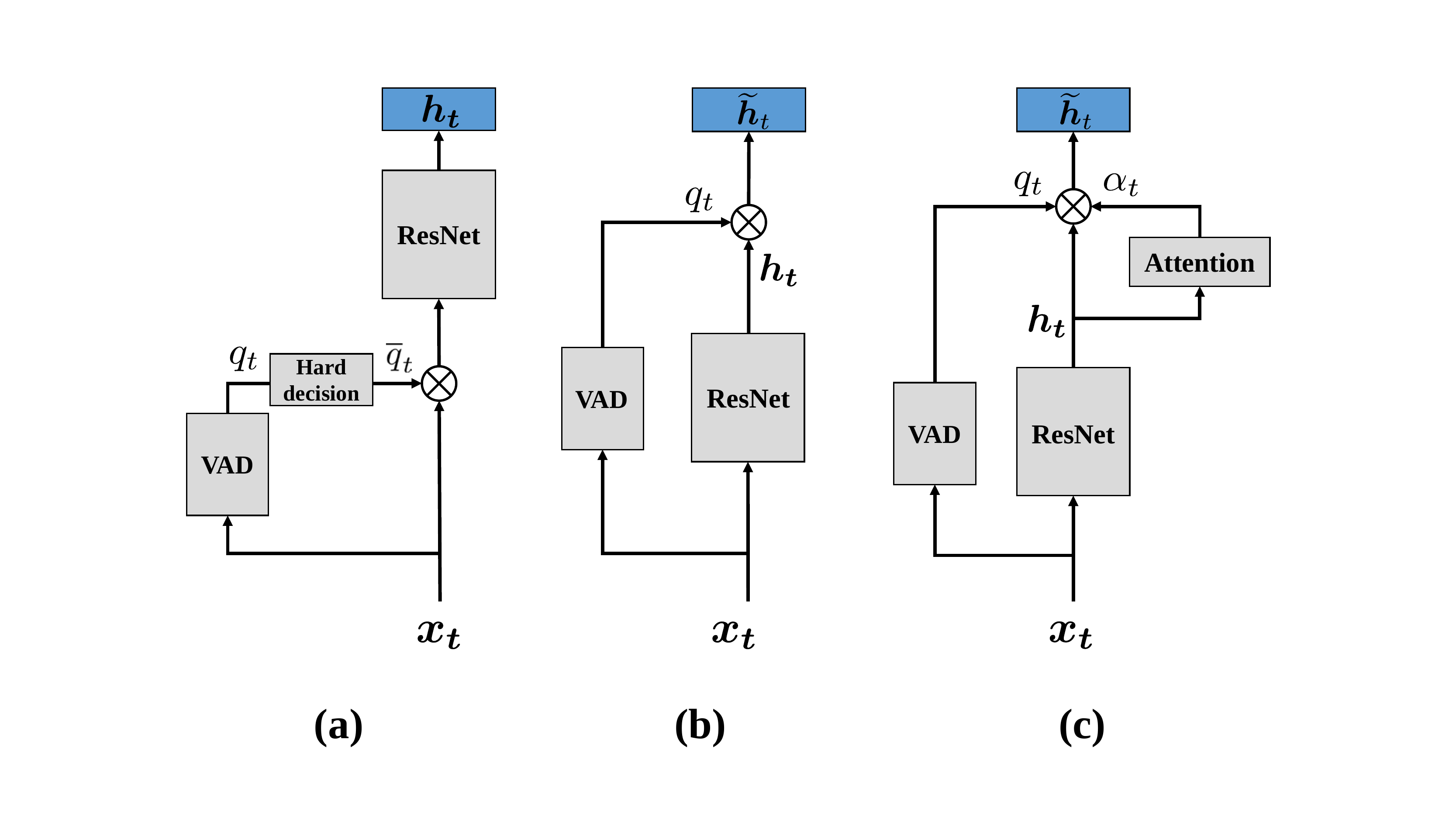}}
  \vspace{-0.4cm}
  \caption{Three combinations of VAD and ResNet-based speaker feature extractor: (a) hard VAD (typical VAD), (b) gating-based soft VAD, (c) attention-based soft VAD.}
  \label{VAD_SR}
\end{figure}

The soft VAD proposed in \cite{Wang2018} is integrated into self-attentive pooling. 
For the acoustic feature $\boldsymbol{x}_t$, the DNN-based VAD produces a frame-wise speech posterior as
\begin{equation} \label{eq:speechposterior}
q_t = \mathcal{VAD}(\boldsymbol{x}_{t-k},\dotsm,\boldsymbol{x}_{t},\dotsm,\boldsymbol{x}_{t+k})\,,
\end{equation}
where $t$ is frame index and $2k+1$ is the context window size, respectively. 
For simplicity, we use the same acoustic features as in the SV system (i.e., 40-dimensional Fbank features with $k$=5).
We combine attention weights with speech posteriors estimated from the DNN-based VAD and replace the attention weight $\alpha_t$ with $\alpha_{t} q_t$ in Eq. (\ref{eq:att_weightedsum}). Therefore, the weighted mean vectors are calculated as follows:
\begin{equation} \label{eq:softSAD}
\boldsymbol{\widetilde{\mu}} = \sum_{t=1}^T \alpha_{t} q_{t} \boldsymbol{h}_{t}\,.
\end{equation}
We call this approach the attention-based soft VAD and this is shown in Fig. \ref{VAD_SR} (c).
Wang \textit{et al.} \cite{Wang2018} showed that applying the speech posterior as a weight in attentive pooling improves the performance in a deep speaker embedding system. 

The third module consists of two fully-connected layers. The first fully-connected layer produces a 128-dimensional speaker embedding $\boldsymbol{y}$. The last layer is a softmax layer, and each of its output nodes corresponds to one speaker ID. 
The model is trained by minimizing the cross-entropy loss over speakers in the training set.
For enrollment, the speaker embedding for each enrollment speaker is stored after length normalization is applied. Finally, scoring between enrollment and test speaker embedding is performed using the cosine distance.

\subsection{DNN-based self-adaptive VAD}
In general, VAD and SV models are trained using different datasets. Therefore, when the VAD is used for SV, the performance of the VAD can be significantly degraded due to the domain mismatch between the source domain (VAD) and target domain (SV) data. 
To reduce the domain mismatch in the VAD, we propose two fine-tuning based unsupervised domain adaptation (DA) methods: speech posterior-based DA (SP-DA) and joint learning-based DA (JL-DA). 

\renewcommand{\algorithmicrequire}{\textbf{Input:}}
\renewcommand{\algorithmicensure}{\textbf{Output:}}
\begin{algorithm}
\caption{Self-adaptive soft voice activity detection}
\label{alg:loop}
\begin{algorithmic}[1]
\Require{Training set $D={\{\mathcal{X}^{v}_i,\mathcal{X}^{s}_i,y^{s}_i}\}_{i=1}^N$, pre-trained VAD model $\mathcal{VAD}_0$, posterior threshold $\delta$, loss weight $\lambda$}
\Ensure{Fine-tuned VAD model $\mathcal{VAD}$ with parameters $\theta_v$, speaker verification model $\mathcal{M}$ with parameters $\theta_s$}
\State $\mathcal{VAD} \gets \mathcal{VAD}_0$
\Repeat{}
    \For{$i := 1$  to  $N$}
    \State // Speech posterior-based domain adaptation
    \State $q_i \gets \varnothing$
        \For{$t := 1$  to  $T_i$}
            \State{$ q_{i,t} \gets \mathcal{VAD}(\boldsymbol{x}_{i,t}^{v};\theta_{v})$}
            \State $q_i \gets q_i \cup \{q_{i,t}\}$
        \EndFor
        \State $\boldsymbol{\hat{X}}^{v}_i, \boldsymbol{\hat{Y}}^{v}_i \gets \mathcal{F}(q_i, \delta)$
        \State $\mathcal{L}_{SP} \gets \mathcal{L}(\mathcal{VAD}(\boldsymbol{\hat{X}}^{v}_i;\theta_{v}), \boldsymbol{\hat{Y}}^{v}_i)$

    \State // Joint learning-based domain adaptation
        \State $ \mathcal{L}_{JL} \gets \mathcal{L}(\mathcal{M}(\boldsymbol{X}^{s}_i;\theta_{s}),y^{s}_i)$
        \State // Calculate losses $\mathcal{L}_{v}$ and $\mathcal{L}_{s}$
        \State $\mathcal{L}_{v} \gets \mathcal{L}_{JL}+\lambda\mathcal{L}_{SP}$
        \State $\mathcal{L}_{s} \gets \mathcal{L}_{JL}$
        \State // Update parameters $\theta_{v}$ and $\theta_{s}$
        \State $\theta_{v} \gets \theta_{v}-\eta_{v} \nabla_{\theta_{v}}\mathcal{L}_{v}$ 
        \State $\theta_{s} \gets \theta_{s}-\eta_{s} \nabla_{\theta_{s}}\mathcal{L}_{s}$ 
    \EndFor
\Until{convergence of $\mathcal{M}$}

\end{algorithmic}
\label{alg:SSVAD}
\end{algorithm}

\subsubsection{Speech posterior-based domain adaptation}
The pseudo-code of the proposed method is given in Algorithm \ref{alg:SSVAD}. 
Suppose we have a dataset $D$ of the SV domain. 
$N$ is the total number of utterances in $D$.
$\mathcal{X}^v_i$ and $\mathcal{X}^s_i$ are a set of acoustic features of the $i$-th utterance for VAD and SV, respectively:
\begin{equation}
\mathcal{X}^v_i = \{\boldsymbol{x}^v_{i,1},\dotsm,\boldsymbol{x}^v_{i,T_i}\}\,,
\end{equation}
\begin{equation}
\mathcal{X}^s_i = \{\boldsymbol{x}^s_{i,1},\dotsm,\boldsymbol{x}^s_{i,T_i}\}\,,
\end{equation}
where $\boldsymbol{x}^v_{i,t}$ and $\boldsymbol{x}^s_{i,t}$ are the $t$-th frame's feature vector in $\mathcal{X}^v_i$ and $\mathcal{X}^s_i$, respectively. $T_i$ is the number of frames in the $i$-th utterance. 
As we already mentioned, we use the same 40-dimensional Fbank features in both tasks for simplicity. 
Here, we only have labels for SV and do not have labels for VAD because, in most cases, it is difficult to obtain VAD labels for SV data. $y^s_i$ is the speaker ID of the $i$-th utterance.

Following this, we obtain a set of speech posteriors $q_i$ from the VAD for all the frames in the $i$-th utterance.
Each speech posterior is compared with the predefined threshold ($\delta$) of 0.7.
If the speech posterior of a frame is larger than the threshold value, it is assumed to correspond to a reliably-labeled speech frame.
On the other hand, if the non-speech posterior of a frame is larger than the threshold value, the frame is regarded as non-speech.
We denote this operation by $\mathcal{F}(q_i, \delta)$ and obtain a set of features $\boldsymbol{\hat{X}}^{v}_i$ and corresponding labels $\boldsymbol{\hat{Y}}^{v}_i$ for VAD.
The VAD model is fine-tuned using the obtained labeled data $\{\boldsymbol{\hat{X}}^{v}_i,\boldsymbol{\hat{Y}}^{v}_i\}$ by minimizing the cross-entropy loss function $\mathcal{L}_{SP}$.
We call this method ``speech posterior-based domain adaptation (SP-DA)".

\subsubsection{Joint learning-based domain adaptation}
First, we integrate the VAD into the SV system using the attention-based soft VAD algorithm which was discussed in Section 3.1. For the $i$-th utterance, the loss $\mathcal{L}_{JL}$ of the speaker embedding model $\mathcal{M}$ is computed using $\boldsymbol{X}^{s}_i$ (a fixed-length segment of 300 frames) and the corresponding label $y^s_i$.
The gradients of $\mathcal{L}_{JL}$ are backpropagated through the VAD and speaker embedding model respectively. 
Since  the  VAD  process  is  partly  guided  by  the loss of the speaker embedding network, the front-end would hopefully be able to produce higher posterior probabilities for frames which are more important for the subsequent SV task.
We call this method ``joint learning-based domain adaptation (JL-DA)".

The self-adaptive VAD is then conducted by combining the two losses as follows:
\begin{equation}\label{eq:twolosses}
\mathcal{L}_{v} = \mathcal{L}_{JL}+\lambda\mathcal{L}_{SP}\,,
\end{equation} 
where $\mathcal{L}_{v}$ is the total loss of the VAD and $\lambda$ is the loss weight for $\mathcal{L}_{SP}$. We denote the combination of the attention-based soft VAD and the DNN-based self-adaptive VAD as the ``self-adaptive soft VAD".

\section{EXPERIMENTAL SETUPS}
\label{sec:typestyle}
\begin{figure}[t]
  \vspace{-0.5cm}
  \centerline{\includegraphics[width=12cm]{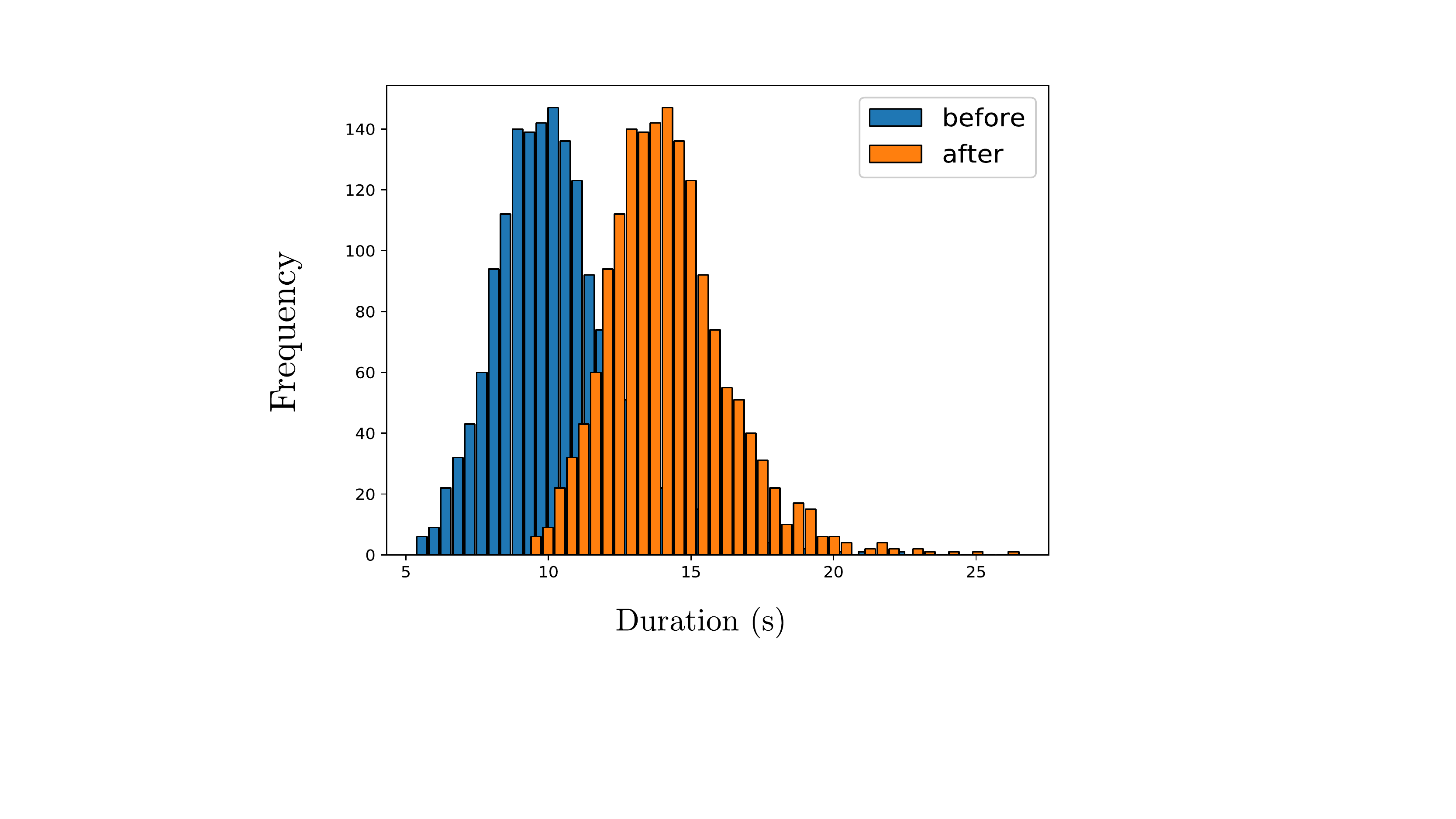}}
  \vspace{-1.6cm}
  \caption{Duration distributions of the test set before and after inserting silence. Mean values are 10 and 14 s, respectively.}
  \label{distribution}
\end{figure}

\subsection{Experimental setups for speaker verification}
We perform experiments on the Korean speech and noise databases \cite{Suh2017}, which were collected by the playback-and-recording method that uses multi-channel microphone arrays for recording the distant speech data. Here, we only use the speech of the first channel among 32 channels.
The speech data was collected at 3 m apart from the artificial mouth 
in an indoor room, which was furnished to simulate a living room acoustically with the reverberation time (RT60) of 0.23 s. 
The noise database consists of 12 types of in-door noise, which was collected using the same approach.
These speech and noise databases are used for creating simulated noisy speech data reflecting various in-door acoustic conditions corrupted by room reverberation and additive noise.

The training set consists of read speech data from 290 speakers and conversational speech data from 260 speakers (550 speakers in total, containing both male and female). For each utterance, the noise is randomly selected from the 3 types of noise (air conditioner, TV, and smartphone ringtone) and added to the distant speech at randomly selected SNRs between 0 and 10 dB, resulting in 200 utterances per speaker.

The utterances of remaining 105 speakers is used for the evaluation.
To simulate more realistic environments where the need for robust VAD is higher, we insert 2 seconds of silence at the beginning and end of the utterance before adding noise. 
The duration distribution of the test data before and after inserting silence are depicted in Fig. \ref{distribution}.
For each utterance, the noise is randomly selected from the 3 types of noise (refrigerator, background conversation, and music) and added to the distant speech at randomly selected SNRs of 0, 5, and 10 dB, resulting in 24 utterances per speaker.
For each speaker, 12 utterances are sampled as the enrollment data. Other than 12 enrolled utterances, we sample 12 utterances each from the same and different speakers. In total, we create 30K trials for testing (in the text-independent scenario). 

The input acoustic features are 40-dimensional Fbank features with a frame-length of 25 ms and a frame shift of 10 ms, which are mean-normalized for each utterance.
All the models are implemented with PyTorch \cite{Paszke2017} and optimized by stochastic gradient descent with momentum 0.9. The mini-batch size is 64, and the weight decay parameter is 0.0001.
We use the same learning rate schedule as in \cite{Jung2019} with the initial learning rate of 0.1.

\subsection{Experimental setups for VAD}
For VAD, we use the same data setup as in \cite{Jung2018}.
To construct the 35 hours training set, the clean training set of the Aurora4 database \cite{Pearce2002} is used. 
To address the class imbalance, 2 seconds of silence is inserted at the beginning and end of the utterance. 
The clean speech is corrupted by 100 types of noise 
at randomly selected SNRs of -5, 0, 5, 10, 15, 20 dB. 

Although we assume that we don't have VAD labels for the SV data, we generate VAD labels to compare the VAD performance before and after domain adaptation. 
We can generate VAD labels because our database has a clean speech corresponding to the noisy speech, unlike most other speaker verification databases.
We apply Sohn VAD \cite{Sohn1999} to the clean speech corpus and the results are used as labels of the corresponding noisy corpus. This method was proved to be sufficiently reasonable to generate VAD labels in \cite{Zhang2016}.
The VAD performance is evaluated with the test data used in speaker verification experiments.

The VAD model has 2 fully-connected hidden layers of 512 units with ReLU activations.
We use the Adam optimizer with a mini-batch size of 512 and the initial learning rate is $10^{-5}$. Self-adaptive soft VAD is fine-tuned by the same optimizer as in SV with the initial learning rate of $10^{-6}$.

\section{RESULTS}
\subsection{Comparison of different VAD methods}


In this section, we compare the results of SV when different VADs are applied. The equal error rates (EERs) are shown in Table \ref{table2}. When we do not use VAD, SV systems using temporal average pooling (TAP) and self-attentive pooling (SAP) show EERs of 13.33\% and 12.31\%, respectively. We observe that energy-based hard VAD (making a hard decision based on a threshold) does not improve the performance of SV under the condition of low SNR with a long silence interval. When we apply DNN-based hard VAD, SV systems with TAP and SAP provide EERs of 11.39\% and 10.83\%, respectively, which are better than when VAD is not applied or energy-based VAD is applied. 

In case domain adaptation (DA) is not used, we compare the hard VAD, G-soft (gating-based soft) VAD, and A-soft (attention-based soft) VAD, which are depicted in Fig. \ref{VAD_SR}. For G-soft VAD, we do not use attention mechanism and only the speech posterior $q_t$ is multiplied to the frame-level feature $\boldsymbol{h}_t$ in Eq. \ref{eq:att_weightedsum}. Here, soft VAD performs better than hard VAD, and A-soft VAD yields a better result than G-soft VAD. 
From this, we can conclude that using the speech posterior and attention weight simultaneously is better than using each one individually. Note that using hard VAD does not improve performance in A-soft VAD, even yielding a higher EER of 10.63\%.

We can see that the SV performance is definitely improved when the self-adaptive VAD and A-soft VAD are used together. When they are combined (i.e., self-adaptive soft VAD is used), we obtain an EER of 9.21\%. 
As a reference, we also provide an upper bound of the performance, which is an EER of 8.27\% obtained by using ground-truth VAD labels.
In case only one of JL-DA and SP-DA is used, the SV system gives a worse result than when they are used together, achieving EERs of 10.75\% and 10.66\%, respectively. 
When only JL-DA is used (i.e., when SP-DA is not used), the VAD is adapted only to minimize the loss of SV rather than being adapted to achieve the original purpose of the VAD task. In this case, the roles of soft VAD and attention mechanism overlap because the attention model is also trained to minimize the loss of SV. 
We can avoid this problem by using the SP-DA together. 
Likewise, we can observe that SP-DA shows better performance by using the JL-DA together. 
We will discuss the reason for this in the following subsection.
\begin{table}[t]
\centering
\caption{Speaker verification EERs using different VADs.} 
\vspace{0.35cm}
\label{table2}
\begin{small}
\begin{tabular}{cccc}
\hline
DA  &  Pooling         & VAD type                         & EER (\%)              \\ \hline
No &   TAP             & No                               & 13.33                 \\
No &   SAP             & No                               & 12.31                 \\
No &   TAP             & Hard VAD (energy)                & 13.35                 \\
No &   TAP             & Hard VAD (DNN)                   & 11.39                 \\
No &   SAP             & Hard VAD (DNN)                   & 10.83                 \\
No &   TAP             & G-soft VAD                       & 10.76                 \\
No &   SAP             & A-soft VAD                       & 10.18                 \\
No &   SAP             & Hard + A-soft VAD                & 10.63                 \\ \hline 
Yes &  SAP             & JL-DA + A-soft VAD               & 10.75                 \\
Yes &  SAP             & SP-DA + A-soft VAD               & 10.66                 \\
Yes &  TAP             & Self-adaptive + G-soft VAD           & 10.59                \\
Yes &  \textbf{SAP}    & \textbf{Self-adaptive + A-soft VAD}  & \textbf{9.21}        \\ \hline
No &   SAP             & Ground-truth VAD labels              & 8.27                 \\\hline
\end{tabular}
\end{small}
\end{table}

\subsection{Effect of the loss weight $\lambda$}
\begin{figure}[t]
  \centerline{\includegraphics[width=11.5cm]{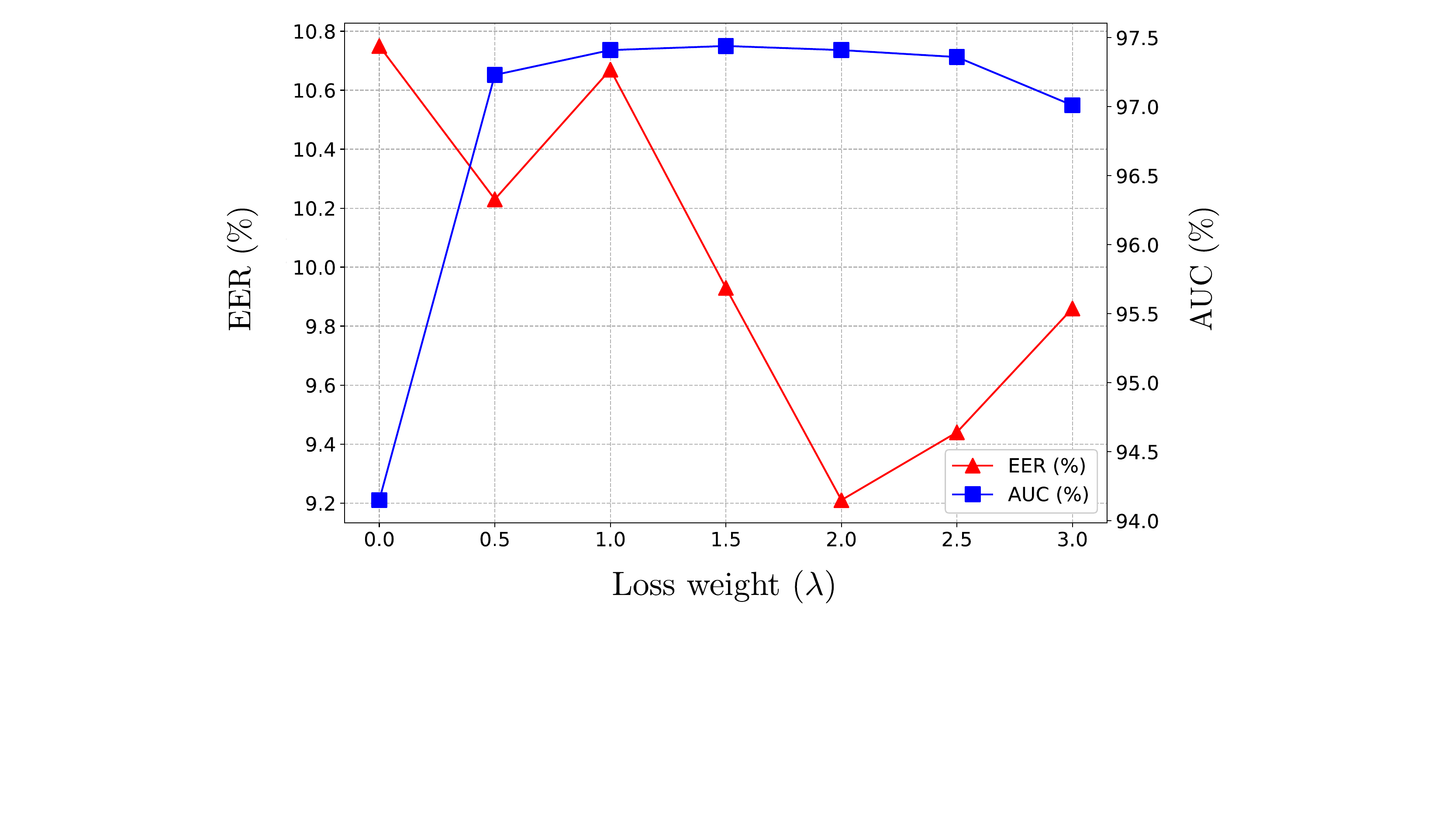}}
  \vspace{-1.8cm}
  \caption{EERs and AUCs with varying the loss weight ($\lambda$).}
  \vspace{-0.1cm}
  \label{lambda}
\end{figure}

Fig. \ref{lambda} shows the performance of SV and VAD as we vary the loss weight $\lambda$ in Eq. \ref{eq:twolosses}. 
As evaluation metrics, we use EER and area under the ROC curve (AUC) \cite{Hanley1982} for SV and VAD, respectively. 
The pre-trained VAD (before domain adaptation) gives an AUC of 91.58\%.

Here, $\lambda$ is the loss weight for $\mathcal{L}_{SP}$ (the loss of SP-DA). 
Therefore, if we increase $\lambda$, then the impact of SP-DA increases. 
When $\lambda$ is 0, we only use JL-DA without SP-DA (i.e., ``JL-DA + A-soft VAD" in Table \ref{table2}). 
In this case, we see that the performance of both SV and VAD is the worst, with an EER of 10.75\% and an AUC of 94.15\%. 
As we already discussed in the previous subsection, this is because the roles of soft VAD and attention mechanism overlap. 
With SP-DA, the VAD can be explicitly adapted to perform its intended purpose which is to classify frames as speech or non-speech.

The AUC value is increasing with the increase of the loss weight $\lambda$ to 1.5. We obtain the highest AUC of 97.44\% at $\lambda$=1.5. However, it can be seen that the AUC value decreases when the $\lambda$ value is greater than 1.5. 
We believe this is because the VAD model is overfitting the adaptation data and losing its ability to generalize. 
If the value of $\lambda$ is extremely large, the influence of SP-DA becomes dominant.
This is the case when we only use SP-DA without JL-DA (i.e., ``SP-DA + A-soft VAD" in table \ref{table2}). In this case, we obtain an EER of 10.66\% and an AUC of 94.98\%. From this point, the AUC value is increasing with the decrease of $\lambda$ to 1.5. 
As a result, we can conclude that JL-DA acts as a regularizer for SP-DA.

According to the figure, EER tends to decrease with increasing AUC, with some exceptions.  
This is consistent with our intuition that the robustness of VAD is directly related to the performance of SV.
The best result is obtained at $\lambda$=2. 
We achieve an EER of 9.21\% with an AUC of 97.41\%.

\section{Conclusions}
\label{sec:foot}

In this paper, we proposed the self-adaptive soft VAD to integrate the DNN-based VAD into the deep speaker embedding-based speaker verification system. 
To reduce the domain mismatch between the VAD and SV data, we combined two algorithms, self-adaptive VAD and soft VAD.
By applying soft VAD into self-attentive pooling, we could fine-tune the VAD to improve the performance of the speaker verification system directly.
Besides, we could obtain VAD labels by thresholding the speech posterior estimated from VAD and fine-tune the VAD using the obtained labeled data.
On the Korean speech dataset, we found that the proposed VAD algorithm outperforms previous approaches for text-independent speaker verification in realistic noisy conditions.
In the future, we will explore how to automatically decide or adapt the threshold in speech posterior-based domain adaptation.

\section{ACKNOWLEDGEMENTS}
This material is based upon work supported by the Ministry of Trade, Industry and Energy (MOTIE, Korea) under Industrial Technology Innovation Program (No.10063424, Development of distant speech recognition and multi-task dialog processing technologies for in-door conversational robots).


\bibliographystyle{IEEEbib}
\bibliography{strings,refs}

\end{document}